\documentclass[11pt]{article}
\usepackage{hyperref}

\begin{document}

\title{Inertia-Dominated Capillary Channel Flow in Microgravity}
\author{Joerg Klatte, Aleksander Grah and Michael Dreyer \\
\\\vspace{6pt}
Zarm - University of Bremen\\
Center of Applied Space Technology and Microgravity\\
Bremen, Germany}
\maketitle

\begin{abstract}
The fluid dynamics video shows three-dimensional experimental and
numerical investigations of open channel flows in microgravity.
The dynamic reorientation of the free surface from 1g to 0g
environment can be observed in a wedge-shaped channel for
subcritical and for supercritical flow rate with a collapse of the
interface. In addition three-dimensional computations determine
important characteristics of the flow, such as the free surface
shape, the velocity field, the dynamics of the reorientation and
the flow rate depended collapse of the free surface. The good
agreement validates the capabilities of the numerical solver.
\end{abstract}

\section{Introduction}
The low gravity environment of the Bremen Drop Tower has been used
to study inertia-dominated open channel flows. The videos
(\href{http://ecommons.library.cornell.edu/bitstream/1813/13724/3/Klatte-MPEG1-LowResolution.mpg}{low-res.},
\href{http://ecommons.library.cornell.edu/bitstream/1813/13724/2/Klatte-MPEG2-HighResolution.mpg}{high-res.})
illustrate two characteristic experiments with subcritical and
supercritical flow rate in a wedge-shaped channel. For subcritical
flow rate the interface can balance the pressure difference
between the gas phase and the liquid. At supercritical flow rate
the decrease of liquid pressure along the flow path due to
convective and viscous momentum transport causes the collapse of
the free surface. Details of the experimental setup are described
in \cite{Haake2009}. The opening angle of the wedge-shaped channel
is $16$ degrees with a channel height of three centimeter. The
experiment liquid is HFE-7500 (3M). The flow rates for the
subcritical case is $2.9$ ml/s, which is close to the critical
flow rate $3.0$ ml/s. The flow rate for the supercritical case is
$5.0$ ml/s. The results are compared to numerical calculations
with the CFD code OpenFoam \cite{OpenFoam}. The computational grid
consists of approx. $200.000$ Hex-elements with user-defined
grading and dynamical refinement at the interface (interDyMFoam).

\end{document}